\documentstyle[preprint,aps]{revtex}
\tightenlines
\begin{document}
\draft
\title{Coulomb glass simulations: Creation of a set of low-energy many-particle
states, non-ergodic effects in the specific heat}
\author{A.\ D\'\i az-S\'anchez$^{1,2}$, A.\ M\"obius$^1$, M.\ Ortu\~no$^2$,
  A.\ P\'erez-Garrido$^2$, and M.\ Schreiber$^3$ \\
 {\it \noindent $^1$Institut f\"ur Festk\"orper- und Werkstofforschung,
  \noindent  D-01171 Dresden, Germany,}\\
 { \it \noindent $^2$Departamento de F\'\i sica, Universidad de Murcia,
  \noindent E-30071 Murcia, Spain}\\
 {\it \noindent \ $^3$Institut f\"ur Physik, Technische Universit\"at Chemnitz,
  \noindent D-09107 Chemnitz, Germany}
}

\maketitle
\begin{abstract}
We have implemented a new numerical method to obtain the low-energy 
many-particle states of the Coulomb glass. First, this method creates an 
initial set of low-energy states by a hybrid of local search and 
simulated annealing approaches. Then, systematically investigating the 
surroundings of the states found, this set is completed. The transition 
rates between these states are calculated. The connectivity of the 
corresponding graph is analysed in dependence on temperature and 
duration of measurement. We study how the formation of clusters is 
reflected in the specific heat as non-ergodic effects.

\end{abstract}

\pacs{65.40.+g, 71.10.-w, 71.55.Jv}

Disordered systems of interacting localized particles have been extensively 
studied for over two decades \cite{PO85,SE84}. A characteristic feature of 
these systems is a complex many valley structure of the energy landscape
of the state space. In semiconductor physics, the Coulomb glass is a 
prominent example. It plays an important role as a semiclassical model 
for a disordered system of localized states with negligible quantum 
tunneling between them.

We consider a half-filled impurity band in the strongly localized
regime represented by the standard tight-binding Coulomb glass 
Hamiltonian \cite{PO85,SE84}:
\begin{equation}
H=\sum_i\epsilon_i n_i +\sum_{i<j} {\frac{(n_i-1/2)(n_j-1/2)}
{r_{ij}}}\,, 
\label{1}
\end{equation}
where $n_i\in \{0,1\}$ denotes the occupation number of site $i$.
The values of the random potential $\epsilon_i$ are uniformly
distributed between $-W/2$ and $W/2$. $r_{ij}$ is the distance between 
sites $i$ and $j$ according to periodic boundary conditions. 
The sites are arranged at random with a minimum separation between 
them, which we choose to be $0.5\,r$, where $r=(4\pi \rho/3)^{-1/3}$, 
and $\rho$ is the concentration of sites. We take $r$ as 
unit of distance, and the Coulomb interaction over a distance 1 as 
unit of energy.

It is a complicated task to obtain the low-energy many-particle states 
\cite{MCPO91,ST93,MOPO96,PO97,TA93}. We have implemented 
a new numerical method which comprises sophisticated local search
\cite{MOPO96,PO97}, thermal cycling \cite{MO97}, and a renormalization 
approach to combinatorial optimization \cite{MODI97}. First, we quench 
states chosen at random by means of a local search procedure, 
ensuring stability with respect to excitations on one up to four sites. 
Thus an initial set of metastable states is created. It is improved by 
cyclically heating (performing a fixed number of successful Metropolis 
steps), and quenching \cite{MO97}. In the course of this process, the 
temperature is decreased stepwise. Finally, we complete the set of 
low-energy states by systematically investigating the surroundings 
of the states found \cite{MOPO96,PO97}. 

The set of low-energy states obtained is the basis for
the study of different low-temperature properties. As a check, we verified 
that our results for the equilibrium specific heat agree with 
\cite{MOPO96,MOTH97}. However, our main aim is to study the influence of 
the duration of measurement, $\tau_m$, on the specific heat values, $c$. 
For that, we have to analyse the rates of the transitions between the 
states.
 
The corresponding transition time (inverse of the transition rate in 
equilibrium) between two many-particle states $I$ 
and $J$ is a product of an energy factor and a spatial factor 
\cite{PO85},
\begin {equation}
\tau_{IJ}=\tau_0 \exp \left( E_{IJ}/kT \right) \exp \left( 2\sum
r_{ij}/a \right)\,,
\label{2}
\end{equation}
where $k$ is the Boltzmann constant, taken as $1$, and $T$ the temperature.
$E_{IJ} = \max (E_I,E_J)$, where the ground state energy
is assumed to be 0. The sum concerns only the sites, which change their 
occupation in the transition; it is the minimized sum of the related 
hopping lengths. $a$ denotes the localization radius, and $\tau_0$ is a 
constant of the order of the inverse phonon frequency, 
$\tau_0\sim 10^{-13}\ {\rm s}$. 

We consider the graph of the transitions between the many-particle 
states: Connections indicate that $\tau_{IJ} <  \tau_m$. Thus the 
states are grouped into clusters. Assuming that thermalisation has 
happened inside the clusters, we measure the specific heat of a 
cluster $\alpha$,
\begin{equation}
c_\alpha =\frac{1}{T^2 \rho} \sum_{I\in \alpha}  
\left( \langle E_I^2 \rangle_\alpha -\langle E_I \rangle^2 _ \alpha
\right)
\label{3}
\end{equation}
where the sum is performed over all states in $\alpha$, and 
$\langle ... \rangle_\alpha$ means thermal average in this cluster.

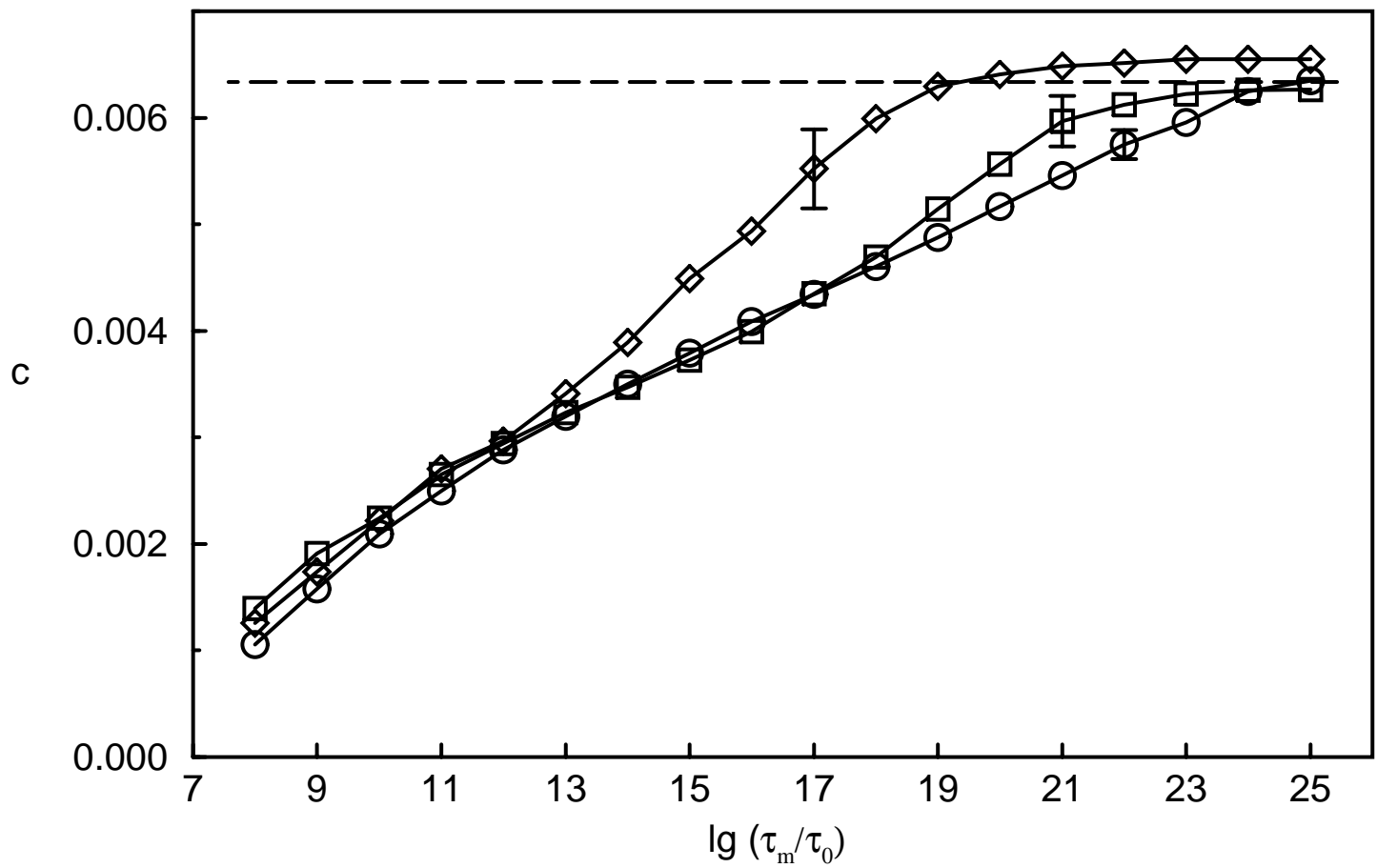
\begin{figure}
\caption{
Dependence of the specific heat, $c$, on the duration of the measurement,
$\tau_m$, for three sizes: $\Diamond $, $\Box $, and $\bigcirc $
denote 64, 216, and 512 sites, respectively. The broken line represents 
the equilibrium value. The localization radius is $a=0.1\ r$, $W=2$, 
$T=0.012$, and $\tau_0\sim 10^{-13}\  {\rm s}$. 
Typical error bars are represented for one point of each curve; the ensemble
averaging took into account 200 `samples'.
}
\label{fig1}
\end{figure}

We obtain the total specific heat, $c(T,\tau_m)$, as weighted average
of the $c_\alpha$. The weight $P_\alpha$ depends on the 
experimental situation simulated. We consider two situations:

A) The `sample' is at equilibrium. Thus $P_\alpha = Z_\alpha /Z$ where 
$Z$ denotes the partition function. Fig.\ \ref{fig1} shows 
$c(T=0.012,\tau_m)$ in comparison to the equilibrium value of $c$, 
where size effect, and reliability region are illustrated. Its main 
result is that also for durations of $1\ {\rm s}$ to several hours 
the specific heat is significantly smaller than the equilibrium value.

B) The `sample' has been quenched from infinite $T$ to the measuring
$T$ within a short time interval. To simulate this we quench first to 
$T=0$, and heat then immediately to the measuring $T$: We start 
assigning the same probability to all states, and connecting them if 
the relaxation time (eq.\ \ref{2} without energy factor) is lower than 
the quenching time $\tau_q$. Note that this graph differs from the
`equilibrium graph'. 

Starting with the highest state of 
the cluster considered, we distribute its weight according to the 
transition probabilities to the states of lower energy. This process 
continues iteratively until only the local minima have a finite 
occupation probability. Finally, we assign to each `equilibrium cluster' 
the sum of the probabilities of the included `non-equilibrium local minima'.

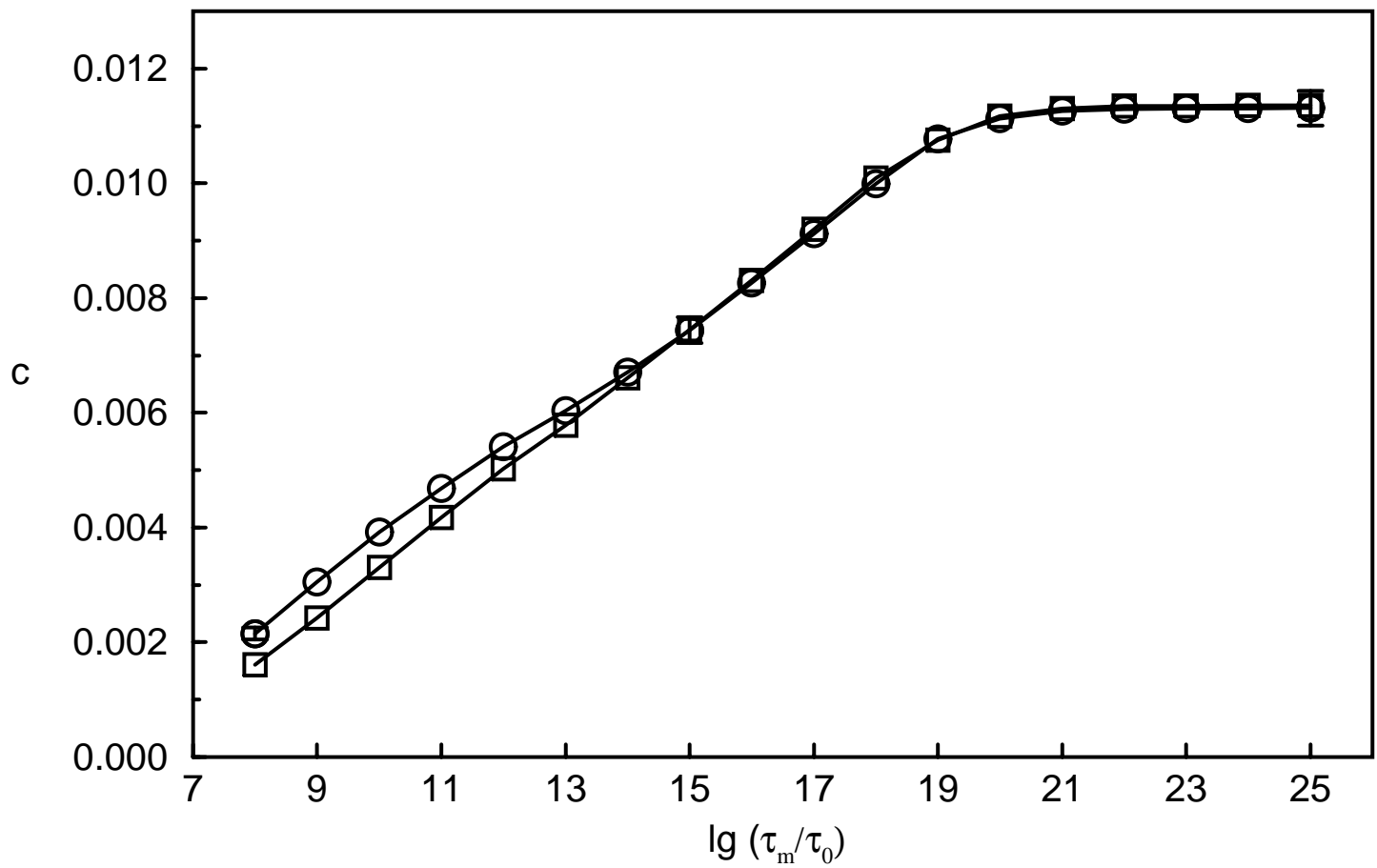
\begin{figure}
\caption{
Comparison of two experiments with different initial conditions: $\bigcirc $ 
= equilibrated `samples', $\Box $ = quenched `samples'. Here $a=0.1\ r$, 
$W=2$, 216 sites, $T=0.018$, and $\tau_q = 10^{12}\  \tau_0 = 0.1\  {\rm s}$. 
Typical error bars are represented for one point of each curve; the ensemble
averaging took into account 200 `samples'.
}
\label{fig2}
\end{figure}

In Fig.\ \ref{fig2} we show the comparisons of both `sample' preparation 
scenarios. Considering `samples' prepared as described in the previous 
paragraph, we obtain almost the same result as for `samples' being in 
equilibrium (case A). Thus the question of the choice of the $P_\alpha$ 
is not an important task for the $\tau_m$ and $\tau_q$ considered.

We would like to acknowledge financial support from the SMWK, and 
from Acciones Integradas HA96-0065.

\end{document}